\documentclass[nohyper,12pt,letterpaper]{JHEP3}

\author{Robert de Mello Koch$^{2,3}$, Antal Jevicki$^1$ and Jo\~ao P. Rodrigues$^2$\\
\qquad \\
Department of Physics$^{1}$,\\ 
Brown University,\\ 
Providence, RI 02912,\\ 
USA\\
\qquad\\
Department of Physics and Centre for Theoretical Physics$^{2}$,\\ 
University of the Witwatersrand,\\ 
Wits, 2050,\\ 
South Africa\\
\qquad\\
Stellenbosch Institute for Advanced Studies$^{3}$,\\
Stellenbosch,\\
South Africa\\
\qquad\\
E-mail: \email{robert@neo.phys.wits.ac.za, antal@het.brown.edu, 
rodriguesj@physics.wits.ac.za}}

\abstract{
Following the recent work of hep-th/0405076 we discuss the emergence of D-brane instanton
solutions in  c=0 noncritical string theory. Our emphasis is on finding the  D-instanton
effects in a field theoretic setting. Using the framework of single matrix collective 
field theory (CSFT) we exhibit the appearance of such solutions.  Some subtle issues regarding 
the form of the field theory equations, the comparison with string equations and the importance 
of a finite N exclusion principle are also discussed.
}

\preprint{}

\title{Instantons in $c=0$ CSFT}

\def \Tr{\mbox{Tr\,}}

\begin{document}

\section{Introduction}
Noncritical $\, c\leq 1\,$ string theory has always been a useful laboratory for 
studying both perturbative and nonperturbative phenomena in string theory[1-6].  Through 
the large $N$ matrix duality nonlinear (string) equations were derived for these 
theories in the early 1990's and the first insight into nonperturbative D-brane phenomena 
was obtained.

Recently the subject has been studied with renewed vigour[7-22]. A highly nontrivial calculation 
was accomplished in ref\cite{kawai} where the D-instanton contribution to the c=0 string 
partition function was evaluated.  The explicit numerical result
\begin{equation}
{i\over 8\sqrt{\pi} 3^{3/4} \, t^{5/8} } \, e^{-{8\sqrt{3}\over 5g_{s}} \, t^{5/4} }
\end{equation}
exhibits the action of the instanton (in the exponential) and the prefactor specifying 
the overall weight 
(chemical potential) of the instanton contribution. The classical action of the instanton 
has been known for some time; it can be evaluated in particular from the above mentioned 
string or loop equations\cite{IK},\cite{JR}. 
It can also be computed as the disk amplitude in conformal Liouville
theory.
The evaluation of the overall coefficient accomplished in ref\cite{kawai} required the 
original, matrix integral 
representation of $\, c=0\,$ string theory. As was emphasized in ref\cite{kawai} this overall 
coefficient 
is not yet computable in conformal (Liouville) theory.  Furthermore attempts to obtain this 
effect (and the instanton solution itself) in the field theoretic formalism of loop equations 
or KdV type string equations have exhibited clear difficulties.  

In the present work, we consider the issue of obtaining the D-instanton effect in field 
theoretic terms.  
We use the framework of collective field theory\cite{JS}
that was successful in various studies of $\,c=1\,$ noncritical string theory.  In the present 
case $(c=0)$ it is convenient to use a stochastic framework, which stabilizes the theory in a manner
similar to the Marinari-Parisi\cite{MP} framework, combined with density function collective field theory.  
Generally, 
we will call this field theoretic representation CSFT.  It 
is similar in form the stochastic approach to loop space string field theory established 
in refs \cite{IK}, \cite{JR}, but with some subtle 
differences which we will exhibit in the text.  Some of these differences do 
play a role in the question of deriving the instanton effect in this version of field theory.

The content of our paper goes as follows.  In Sect.2 we discuss some basic facts involved in 
the
collective integral representation. In particular the role of the Jacobian is emphasized. We 
explain that the presence of this Jacobian plays a crucial role in supplying the correct 
weight in the
instanton sector. In Sect.3 we describe the stochastic (Fokker-Planck) version of collective
field theory. We comment on the similarities (and differences) in comparison to loop space 
field theory and discuss how the string equations emerge in the scaling limit.  In Sect.4 we 
exhibit and solve the equations characterizing the single eigenvalue instanton. Sect.5 is 
reserved for conclusions.

\section{Basics}

In this section, we begin by outlining some (well known) basic elements of collective field 
theory.  One starts with the c=0 partition function in the 
eigenvalue representation
\begin{equation}
Z_N = \int \prod_{i=1}^N \, dx_i \Delta^2 \, e^{-\sum_{i} V (x_{i} )}.
\end{equation}
The field theoretic representation is achieved by changing to the collective field 
(density of eigenvalues)
\begin{equation}
\rho (x) = \sum_{i=1}^N \, \delta (x-x_i ).
\end{equation}
The potential with the van der Monde measure gives rise to the action
\begin{equation}
S(\rho ) = - {1\over 2} \int \,dx\, dy\,\rho (x) \ln (x-y)^2 \rho (y) 
+ \int \,dx\, \rho (x) V (x).
\end{equation}
The partition function is now
\begin{equation}
Z_N = \int \prod_x d\rho (x) \quad J_N (\rho ) e^{-S(\rho )}
\end{equation}
with a nontrivial Jacobian $\, J_N (\rho )\,$ arising through the change of variable
\begin{equation}
J_N (\rho ) = \int \prod_{i=1}^N \, dx_i \prod_x \delta \left( \rho (x) - \sum_{i=1}^N \delta 
(x - x_i )\right).
\end{equation}
It is this Jacobian (which is usually ignored) that will play some role in the field theoretic 
treatment of the eigenvalue instanton.

The Jacobian is nontrivial both in its nonpolynomial functional dependence on the collective 
density
and its nontrivial scaling properties (with respect to $N$).
An explicit (power series) expansion for the Jacobian as function of $\, \rho (x)\,$ can be
generated as follows.  Using a Lagrange multiplier $\,\psi (x)\,$ one has

\begin{equation}
J_N (\rho ) = \int [ d\psi ] \, e^{i\int \psi(x)\rho(x) dx} \, 
\left[ {1\over L} \int e^{-i\psi(x')}dx' \right]^N.
\end{equation}
In the limit $\, N\rightarrow \infty$, $\, L\rightarrow \infty :  \rho_0 = N/L\,$ we have
\begin{equation}
\lim_{N,L\to\infty} \left( {1\over L} \int dx \, e^{-i\psi (x)} \right)^N 
= e^{\rho_{0} \int dx \left[ e^{i\psi (x)} \right]_{ir}}
\end{equation}
where $\,ir\,$ denotes the irreducible part.
This representation implies that the collective field theory can be written in terms of 
the two coupled field $\, \rho (x)\,$ and $\, \psi (x)\,$ with the action
\begin{equation}
S [\rho , \psi ] = \int dx \left( i\psi (x) \rho (x) + {N\over L} \left[e^{-i\psi (x) } 
\right] \right) + S (\rho )
\end{equation}
and with a trivial measure i.e. no Jacobian. Since $\, \rho (x)\,$ appears quadratically in 
the above, it can also be eliminated resulting in a Lagrangian for $\, \psi (x)\,$ only. We 
will not use this ``dual" representation in this work and will not pursue it further.

The Jacobian enforces several nontrivial features contained in the above transformation.  
First one has an infinite chain of constraints.  These can be described in terms of the 
moments $\, \rho_n = \int dx \, x^n \rho(x)\,$ as follows:  introduce the Schur polynomials, 
ref\cite{Jev}
\begin{equation}
P_n \left( \rho_1 , \rho_2 , \rho_3 ,\cdots\right).
\end{equation}
The constraints can be written as
\begin{equation}
P_{N+n} \left( \rho_1 , \rho_2 , \rho_3 \cdots \right) = 0.
\end{equation}
They imply the fact that the variables $\, \rho_{N+n}\,$ are dependent on 
$\, \rho_1 , \rho_2 , \cdots \rho_{N-1} $.
We see that the Jacobian enforces the exclusion principle which manifests itself as a cut off 
at $\, n=N\,$. The second property which can be seen to follow from the Jacobian is a 
recursion 
involving $\, N\rightarrow N-1$.  This will be of crucial importance for the instantons.

Let us then consider the one-instanton sector obtained by separating a single eigenvalue. 
Write
\begin{equation}
\rho (x) = \rho' (x) + \delta (x-y)
\end{equation}
where for $\, \rho' (x) : \int dx \rho' (x) = N-1$. In this case
\begin{equation}
\int [d\rho ] \, J_N (\rho + \delta ) = N \int dy \int [d\rho' ] \, J_{N-1} (\rho' ).
\end{equation}
Consequently
\begin{equation}
Z_N = Z_N^{(0)} + N \int dy \int [d\rho' ] \, J_{N-1} (\rho' )\, e^{-S (\rho' + \delta )}
+...
\end{equation}
where $\, Z_N^{(0)}\,$ denotes the no instanton sector partition function.  Here we 
see that the recursion property of the Jacobian supplies a crucial 
factor of $\, N\,$ in the weight of the $\, 1\,$ instanton contribution.
The integral over the location of the instanton $\, y \,$ is a standard field theory 
collective coordinate integration.

The rest of the calculation now proceeds by evaluating the 1-instanton sector functional
integral through the stationary point method.  The action becomes

\begin{equation}
S (\rho + \delta ) = - {1\over 2} \int \rho'(x) \ln (x - x')^2 \rho'(x') \,dx\,dx'\,
+ \int V (x) \rho' (x)\, dx\, - 2 \int \rho' (x) \ln (y-x) \, dx\,
\end{equation}
where
\begin{equation}
\int dx \rho' (x) = N-1 = N'.
\end{equation}
Rescaling $\rho'\,$ and $\, x\,$ we have
\begin{equation}
S /N^{\prime 2} = - {1\over 2} \int \rho(x) \ln (x - x' )^2 \rho (x') \,dx\,dx'\,
+ \int V(x) \rho\,dx\, - {2\over N'} \int \rho \ln (y-x)\,dx\,
\end{equation}
and for the rest of the discussion that follows we will have that $\, \rho(x)\,$ is normalized 
to 
$\, 1: \int dx \rho (x) = 1 $. The equations of motion following from this (shifted) action 
are
\begin{equation}
2\int dx'\rho(x')\log (x-x')=V(x)-{2\over N'}\ln (y-x).
\end{equation}
The derivative with respect to $x$ of this equation takes the form of a BIPZ one matrix
integral equation\cite{BIPZ}
\begin{equation}
2\int \!\!\!\!\!\!\! - \,\,\,\,\,\, 
dx'\rho(x'){1\over x-x'}=V'(x)+{2\over N'} {1\over y-x}\equiv V'(x)+{1\over N'}\Delta V'.
\end{equation}
This equation can be exactly solved using the standard methods as will be described in 
appendix A. In this section, we will
adopt a field theoretic approach allowing a construction of $\rho(x)$ to first order in 
${1\over N'}$ since we see that the eigenvalue term in the equation represents a source with a 
coupling proportional to 1/N . Therefore one can expand
\begin{equation}
\rho(x)=\rho_0(x)+{1\over N'}\rho_1(x)+O\left({1\over N^{\prime 2}}\right)
\end{equation}
where the leading order $\rho_0(x)$ represents a solution to the equation

\begin{equation}
\int \!\!\!\!\!\!\! - \,\,\,\,\,\, {\rho_0(x')\over x-x'}dx'={1\over 2}V'(x).
\end{equation}

\noindent
The correction $\rho_1(x)$ can be expressed in terms of the density Greens function\cite{BZ}
which reads

\begin{equation}
K_2(x,x')={1\over 2\pi^2}{1\over (x-x')^2}{xx'-{1\over 2}(a_0+b_0)(x+x')+a_0b_0\over
\sqrt{(x-a_0)(x-b_0)(x'-a_0)(x'-b_0)}}
\end{equation}

\noindent
as

\begin{equation}
\rho_1(x)=\int dx'K_2(x,x')\Delta V(x')=-2\int dx'K_2(x,x')\log (y-x').
\end{equation}

\noindent
To perform the above integral, it is useful to write the Greens function as

\begin{equation}
K_2 ={1\over 2\pi^2}{1\over \sqrt{a_0-x}\sqrt{x-b_0}}{\partial\over\partial x'}
{\partial\over\partial x}\left(\sqrt{a_0-x'}\sqrt{x'-b_0}\log |x-x'|\right).
\end{equation}

\noindent
After an integration by parts, we obtain

\begin{equation}
\rho_1(x)={1\over \sqrt{a_0-x}\sqrt{x-b_0}}{1\over y-x}
\int^{a_0}_{b_0} {dx'\over \pi^2}\sqrt{a_0-x'}\sqrt{x'-b_0}\left[
{1\over x-x'}-{1\over y-x'}\right].
\end{equation}

\noindent
Next consider the computation of integrals of the form

\begin{equation} 
I=\int^{a_0}_{b_0} dx'\sqrt{a_0-x'}\sqrt{x'-b_0}{1\over x-x'}.
\end{equation}

\noindent
After a shift in the integration variable, this integral reduces to 
computing the Hilbert transform

\begin{equation}
{1\over\pi}\int_{-a}^a {\sqrt{a^2-x^2}\over \lambda-x}dx=\lambda-\sqrt{\lambda^2-a^2},
\end{equation}

\noindent
for $a<\lambda$ and

\begin{equation}
{1\over\pi}\int_{-a}^a {\sqrt{a^2-x^2}\over \lambda-x}dx=\lambda,
\end{equation}

\noindent
for $a>\lambda$. Thus,

\begin{equation}
\rho_1(x)=-{1\over \pi}{1\over y-x}{\sqrt{y-a_0}\sqrt{y-b_0}\over\sqrt{a_0-x}\sqrt{x-b_0}}
+{1\over \pi}{1\over\sqrt{a_0-x}\sqrt{x-b_0}}.
\end{equation}

\noindent
Regarding this expression we note that the coefficient of ${1\over y-x}$ can be seen 
to match with the exact result given in Appendix A.

It should be noted that although in this paper we concentrate on the $c=0$ criticality, 
this result is universal and independent of the potential, as it was obtained from the
universal correlator of \cite{BZ} together with the logarithmic source associated with the
instanton eigenvalue. 

We are now in a position to evaluate the partition function in the one instanton 
sector. We have already seen that in this case the partition function takes the form

\begin{equation}
Z_N^{(1)}=N \int dy \int [d\rho' ] \, J_{N-1} (\rho' )\, e^{-S (\rho' + \delta )}.
\end{equation}
Evaluating this through the stationary point method gives 
\begin{eqnarray}
Z_N^{(1)}&=&N \int_{\rm out} dy \, e^{-S^1 (\rho_0 + {1\over N'}\rho_1 )}\nonumber\\
&=&Z_{N-1}^{(0)}N\int_{\rm out} dy e^{2N'\int \rho_0(x) \ln (y-x) dx 
+2\int \rho_1(x) \ln (y-x) dx}.
\end{eqnarray}
where we have identified the no instanton partition function
\begin{equation}
Z_{N-1}^{(0)}=e^{-N^{\prime 2}S(\rho_0 )}
\end{equation}
for the $N-1\times N-1$ matrix model. The label ``out" for the $y$ integration region 
indicates that we are integrating out of the region of $\rho_0$'s support, that is 
$y>a_0$ and $y<b_0$. The $O(1)$ piece of the integrand
\begin{equation}
e^{2N'\int \rho_0(x) \ln (y-x) dx + 2\int \rho_1(x) \ln (y-x) dx}
\end{equation}
is independent of the potential. Evaluating this contribution we obtain
\begin{eqnarray}
2\int \rho_1(x) \ln (y-x) dx &=&4 \int \log (y-x')K_2(x',x)\log (y-x)\nonumber\\
&=& 2\log\left[ 1+{y-{a_0+b_0\over 2}\over\sqrt{y-a_0}\sqrt{y-b_0}}\right]
-2\log 2,
\end{eqnarray}
where we see agreement with \cite{kawai}. Continuing, the complete contribution 
of the 1-instanton normalized with respect to the no instanton (vacuum) partition function 
is given by 
\begin{equation}
\mu={Z_N^{(1)}\over Z_N^{(0)}}={N\over 4}\int_{\rm out} dy 
{Z_{N-1}^{(0)}\over Z_N^{(0)}}
\left[ 1+{y-{a_0+b_0\over 2}\over\sqrt{y-a_0}\sqrt{y-b_0}}\right]^2 
e^{-NV(y)-2N\int \rho_0(x) \ln (y-x) dx}.
\label{formu}
\end{equation}
The remaining integral over $y$ is evaluated through the stationary point method.
Here for the ratio of partition functions we have 
\begin{equation}
{Z_{N-1}^{(0)}\over Z_N^{(0)}}={e^{-N^{\prime 2}S(\rho_0 )}\over e^{-N^{2}S(\rho_0 )}}
=e^{(2N-1)S(\rho_0 )}.
\end{equation}
Notice that it is the leading order free energy that determines this ratio.
The equation (\ref{formu}) for $\mu$ 
above represents a general expression for a single eigenvalue contribution,  
written in terms of a 
generic potential $V(x)$ and can be used to take the scaling limit. As emphasized by 
ref\cite{kawai} the presence of the large $N$ factor $N$ is crucial for obtaining the 
correct weight (chemical potential) of the instanton. In our derivation the factor of 
$N$ is seen to be associated with the field theoretic measure contained in the Jacobian  
$J (\rho) $. In turn any field theory not containing such a measure will not be able to 
provide a correct (i.e. in agreement with a matrix model) prediction of the instanton effect.

\section{Fokker-Planck Collective Field Theory}

The most natural scheme for deriving elements of string field theory from matrix
models is through stochastic quantization. Stochastic (and also the Marinari-Parisi) formulation 
of the one matrix integral provides a stabilization of the model and was as such
originally introduced in \cite{halpern},\cite{karmigdal}.

For the single matrix problem, for a general action $S = \Tr V(M)$, one has
the Fokker-Planck Hamiltonian
\begin{equation}
H_{FP} = - {1\over 2} \Tr \, \left( {\partial\over\partial M} - {\partial S 
\over \partial M} \right) {\partial\over \partial M}
\end{equation}
whose Hermitian form is
\begin{equation}
\hat{H} = - {1\over 2} \left[ {\partial\over \partial M} - {1\over 2} \, 
{\partial S\over \partial M} \right] \left[ {\partial\over\partial M} + {1\over 
2} {\partial S \over \partial M} \right] .
\end{equation}
This equals
\begin{equation}
\hat{H} = - {1\over 2} \left( {\partial^2\over \partial M^2} \right) + {1\over 
4} \left[ S^{(1)} (M) , P\right] + {1\over 8} \Tr \left( S^{(1)} 
(M)^2\right)
\end{equation}
with $P = \partial /\partial M$ and $S^{(1)} (M)= \partial S/\partial M$.  
A similar form is also associated with the Marinari-Parisi supersymmetric
one dimensional matrix model\cite{MP}.  
A heuristic  way to obtain a collective field representation for 
this theory is by the replacement
\begin{eqnarray}
M &\rightarrow& \lambda_i\\
P &\rightarrow& - {(1-\delta_{ij})\over (\lambda_i - \lambda_j )}
\end{eqnarray}
so that
\begin{equation}
\Tr \left[ S^{(1)} (M) , P\right] = - \sum_{i\not= j} \, {v^{(1)} (\lambda_i )   
- v^{(1)} (\lambda_j )\over\lambda_i - \lambda_j}.
\end{equation}
In terms of the density variable
\begin{equation}
\phi (x) = \sum_i \, \delta (x - \lambda_i )
\end{equation}
one has
\begin{eqnarray}
H_{coll} &=&\int {1\over 2} \left[\phi \Pi_{,x}^2 + 
{\pi^2\over 3} \phi^3 + {1\over 8} \left( v^{(1)} (x)\right)^2 \phi (x) 
\right]dx\nonumber\\
&-& {1\over 4} \int dx dy \, {v^{(1)} (x) - v^{(1)} (y)\over x-y}   \,  \,  \phi 
(x) \phi (y).
\end{eqnarray}
This field theory was introduced first in connection with a {\bf supersymmetric } 
generalization of ordinary collective field theory\cite{susycft} and was studied in more 
detail in \cite{feinberg}.  A more rigorous derivation of the extra terms will be 
given latter. This hermitian Hamiltonian can be written as 
\begin{equation}
H = \int {1\over 2} \phi \Pi_{,x}^2\, dx + \int {1\over 2} \phi(x) \left( {-\mskip-19mu\int}  
\, {1\over x-y} \, \phi (y)\,dy - {v' (x)\over 2} \right)^2 dx. 
\end{equation}
Its static stationary point equation is
\begin{equation}
{-\mskip-19mu\int} {\phi (y)\over x-y}dy  = {v'(x)\over 2}
\end{equation}
which is the BIPZ one matrix integral equation. We mention that the above, nonlocal
form for the collective hamiltonian is also used in studies of extremal, BPS soliton
type solutions ref\cite{andric}.

Because of later relevance we comment at this point that the structure involved in the
above hamiltonian is also involved in the Schwinger-Dyson approach which is based on 
using loop variables $\phi_n = \Tr (M^n )$ $n> 0$ or 
$\phi (\ell ) = \Tr (e^{-\ell M} ) \, \ell \geq 0$.  Equivalently one has 
the resolvent

\begin{equation}
\Phi (z) = \Tr \left( {1\over z-M}\right).
\end{equation}
In comparison the collective density field equals

\begin{equation}
\phi (x) = \Tr \left( \delta (x-M)\right) = \int {dk\over 2\pi} \, e^{ikx} \, 
\Tr \left( e^{-ikM} \right)
\end{equation}
with $k \mathop<\limits_{>}0$. Consequently we can think of this as
{\bf also} adding negative loops $\phi (-l ) = \Tr (e^{lM})$ , then
after an analytic continuation $(l\rightarrow ik)$ one obtains
\begin{equation}
\phi (k) = \Tr (e^{ikM} ) \qquad - \infty < k < + \infty .
\end{equation}
We now proceed to the more detailed discussion of how a hermitian collective field 
hamiltonian is derived. One has first that
\begin{equation}
H = - \int_{-\infty}^{\infty} \, dk \left[  \int_{-\infty}^{\infty}  dk' \, 
\Omega (k,k')  \Pi_{k'} + \omega(k) -  \Omega (S, k)  \right]  k \Pi (k)
\end{equation}
with

\begin{eqnarray}
\Pi (k) &=& {\delta\over\delta\phi (k) }\\
\Omega (k,k' ; \phi )  &=& {\partial\phi_k\over \partial M} \, 
{\partial\phi_{k'}\over \partial M} = -k k' \phi (k + k' )\\
\omega (k) &=& \Tr\partial_M^2 \phi (k) = - k^2 \int_0^1 d \alpha \phi (k\alpha 
) \phi \left( k (1-\alpha )\right)
\end{eqnarray}
and
\begin{equation}
\Omega (S, k ) = \Tr {\partial S\over \partial M} \, {\partial\phi_{(k)}\over 
\partial M} = \int_{-\infty}^{\infty} \Omega (k,k') \, {\delta S\over \delta 
\phi_{k'}} \, dk'.
\end{equation}
For comparison, the loop space field theory would also involve the operator
\begin{equation}
O_{\ell} = \phi_{\ell +\ell '} \ell ' \Pi_{\ell '} + \phi_{\ell - \ell '} 
\phi_{\ell '} - \phi_{\ell + \ell '} \ell ' {\delta V\over\delta\phi_{\ell '}}.
\end{equation}
It then involves a shift 
\begin{equation}
\Phi (z) = {1\over 2} V^{(1)} (z) + \varphi (z)
\end{equation}
after which $\varphi$ is taken to continuum limit.  In principle this 
can be  problematic.   
Recall that $\Phi (z) = \sum_{n \geq 0} \, z^{-n-1} \phi_n$ while
$V(z) = \sum \, t_n z^n$ (positive powers).  Consequently one is attempting to 
cancel positive powers through a shift of a variable containing only negative 
ones.  Similarly in loop notation the quadratic term
\begin{equation}
\int_0^{\ell} \ell ' \phi_{\ell - \ell '} \phi_{\ell '}
\end{equation}
involves loops of length $\leq \ell$ while the linear term
\begin{equation}
\phi_{\ell + \ell '} \, \ell ' \, V_{\ell '}^{(1)}
\end{equation}
contains a loop of length $> \ell $. 

Some of these issues are not there in  the density collective representation
which involves both positive and negative loops. The density variable being real and since the 
F-P hamiltonian was hermitian the next (important) step in the collective formalism 
corresponds 
to a similarity transformation which provides a manifestly hermitian representation for 
the Hamiltonian 
\begin{equation}
H = {1\over 2}  \int_{-\infty}^{\infty}  dk\int_{-\infty}^{\infty}  dk'
\, \tilde{O}_k \, \Omega^{-1} (k,k' ) \, \tilde{O}_{k'}^+ 
\end{equation}
with
\begin{eqnarray}
O_k &=& \Omega (k, k') k' \Pi_{k'} + {1\over 2} \left( \omega (k) - 
\Omega (S,k)\right)\\
O_k^+ &=& -k \Pi_{k'} \Omega (k, k') + {1\over 2} \left( \omega (k) - \Omega 
(S,k)\right).
\end{eqnarray}
Apart from ordering terms this gives
\begin{equation}
H = \int dx {1\over 2} \phi \, \Pi_{,x}^2 + V_{eff}
\end{equation}
with
\begin{equation} 
V_{eff} = {1\over 8} \left[ \omega (k) - \Omega (s,k)\right] \Omega^{-1} 
(k,k') \left[ \omega (k') - \Omega (S,k')\right].
\end{equation}
Using

\begin{eqnarray} 
\Omega (x,y) &=& \partial_x \partial_y \left( \delta (x-y) \phi 
(x)\right)\\
\omega (x) &=& 2 \partial_x \left( \phi {-\mskip-19mu\int} {1\over x-y} \phi (y) 
dy\right)
\end{eqnarray}
we obtain the total Hamiltonian
written as

\begin{equation}
H={1\over 2}\int dx\left[\phi\Pi_{,x}^2+\phi(x)\left(
\int \!\!\!\!\!\!\! - \,\,\,\,\,\, {\phi(x')\over x-x'}dx'
-{1\over 2}v'(x)\right)^2\right].
\end{equation}
The effective potential term in this Hamiltonian involves the Hilbert transform and as 
such looks non-local.
Using a certain cubic identity, it can be also be written in the
manifestly local form
\begin{equation} 
H =\int {1\over 2} \left[\left( \phi \Pi_{,x}^2 + 
{\pi^2\over 3} \phi^3 \right) + {1\over 8} \left( v'(x)\right)^2 \phi (x) 
\right]dx
- {1\over 4} \int\, dx dy \, {v'(x) - v' (y)\over x-y}\,\phi 
(x) \phi (y).
\end{equation}

Regarding the two different versions  of the field theory Hamiltonian
we note that their equivalence is based on a formal identity.  In particular,
this might only be strictly true  when there are no divergences (in the energy). In general
the term involving the original Hilbert transform and the local cubic term do not 
necessarily
regularize the divergences in the same way\cite{Andre}. This issue, we believe, is similar to
the issue of surface terms in a typical theory of gravity. As is well known, in any gauge 
fixed version of gravity, surface terms have to be carefully adjusted.
In the above, this observation is of particular relevance for the appearance (or nonappearance) 
of 
instanton solutions. As we will see, the separation of single eigenvalue instantons is possible 
in
the first version of the theory but very questionable in the second.

The second local version where the interaction is given by a simple local cubic term 
corresponds to
the dynamical Fermi surface picture which provided great insight in studies of the c=1 string
theory ref\cite{2d}.  This Fermi surface is also very useful for making contact with the 
scaling string equations. Take for example the potential relevant for the simplest k=2 case:

\begin{equation} 
v(x)={x^2\over 2}-{g\over 3}x^3 ,
\end{equation}

\noindent
then

\begin{equation}
v'(x)=x-gx^2 
\end{equation}

\noindent
and the extra term reads

\begin{equation}
-{1\over 4}\int dx \int dy \big[1-g(x-y)\big]\phi(x)\phi(y)=-{1\over 4}
\left[\left(\int dx\phi\right)-2g\left(\int\phi dy\right)\left(\int dx x
\phi(x)\right)\right].
\end{equation}

\noindent
Normalizing $\int\phi =1$ we have

\begin{equation}
V_{eff}={1\over 2}\left({\pi^2\over 3}\phi^3(x)+\left[
gx+{1\over 4}(x-gx^2)^2\right]+c_0\right),
\end{equation}

\noindent
with the static equation

\begin{equation}
\pi^2\phi^2(x)+\left( gx-c+{1\over 4}(x-gx^2)^2\right)=0.
\end{equation}

\noindent
The double scaling limit is taken as before

\begin{eqnarray}
x=x^* -\gamma ay\\
g=g^* (1-{\sqrt{3}\over 16}\gamma^2 a^2 t)
\end{eqnarray}

\noindent
with

\begin{equation} 
g^*=3^{1/4}/6,\qquad x^*=(\sqrt{3}+1)3^{1/4}
\end{equation}

\noindent
to give (to obtain this equation we have set $\gamma =2^{2/3}3^{5/12}$)

\begin{equation}
(\pi\tilde{\phi})^2-(y^3-{3\over 4}\Lambda y+{1\over 4}\Lambda^{3/2})=0.
\end{equation}

\noindent
Continuing to the hamiltonian we have that the time of flight

\begin{equation}
T=-\int {dy\over\tilde{\phi}(y)}\approx {\epsilon^{2/3}\epsilon}
\left({2\sqrt{2\omega\over 3}}\right)\int {dx\over\pi\phi_0(x)}
\end{equation}

\noindent
is finite 

\begin{equation}
T=\lim_{\epsilon\to 0}\epsilon^{1/2}{1\over \sqrt{\epsilon}}=O(1),
\end{equation}

\noindent
and that the  Hamiltonian becomes

\begin{equation}
H=\epsilon^{1/2}\int dy\left({1\over 2\kappa}(\partial\eta)\pi^2+{1\over 6\kappa}
(\partial\eta )^3+{1\over 2}\phi_0(y)\left(\pi^2+(\partial_y\eta )^2\right)\right).
\end{equation}

Consequently after rescaling we have a finite Hamiltonian which takes the form of a
dynamical Fermi surface theory:

\begin{eqnarray}
{\cal H}&=&\int {dy\over 2\pi}\int_{\alpha_-}^{\alpha^+}dp(p^2 -p_0(y)^2))\\
\alpha_{\pm}\Pi_{,y}&\pm&\pi\tilde{\phi}(y)
\end{eqnarray}

\noindent
with the stationary equation

\begin{equation}
p^2-p_0(y)^2=0.
\end{equation}

\noindent
As a final comparison we mention that the representation obtained can be directly 
compared \cite{yoneya} to the  string equation

\begin{equation}
\left[P,Q\right]=1.
\end{equation}

\noindent
which is based on the Kdv data

\begin{eqnarray}
Q&=&d^2-u\\
P&=&\left(Q^{2k-1\over 2}\right)_+ .
\end{eqnarray}

For $k=2$ one has

\begin{equation}
P=d^3-{3\over 4}\{ u,d\} ,
\end{equation}

\noindent
and in the semiclassical tree approximation

\begin{equation}
P^2=d^6-3ud^4+{9\over 4}u^2 d^2 .
\end{equation}

\noindent
Using $Q=d^2-u$ one has

\begin{equation}
P^2=Q^3 -{3\over 4}uQ+{1\over 4}u^3 .
\end{equation}

\noindent
This is indeed identical to the stationary collective field equation established above.
A precise correspondence reads

\begin{eqnarray}
P&\leftrightarrow&\pi\tilde{\phi}\\
Q&\leftrightarrow& y
\end{eqnarray}
and $u=\Lambda(=t^{1/2})$ from the string equation. 

\section{Instanton Solution in F-P Field Theory}

In this section we discuss how the instanton solution appears in
Fokker-Planck collective field theory. Towards this end, separate out an 
eigenvalue

\begin{equation}
\phi (x)=\tilde{\phi}(x)+{1\over N}\delta (x-y).
\end{equation}

\noindent
The field theory potential derived in the previous section splits into two contributions

\begin{eqnarray}
V_1&=&{1\over 2N}\int dx\delta (x-y)\left(
\int \!\!\!\!\!\!\! - \,\,\,\,\,\, {\phi(x')\over x-x'}dx'-{1\over 2}v'(x)\right)^2
\nonumber\\
&=&{1\over 2N}\left(
\int \!\!\!\!\!\!\! - \,\,\,\,\,\, {\phi(x')\over y-x'}dx'-{1\over 2}v'(y)\right)^2 ,
\end{eqnarray}

\noindent
and

\begin{eqnarray}
V_2&=&{1\over 2}\int dx\tilde{\phi}(x)
\left(
\int \!\!\!\!\!\!\! - \,\,\,\,\,\, {\tilde{\phi}(x')\over x-x'}dx'
+{1\over N(x-y)}-{1\over 2}v'(x)\right)^2\nonumber\\
&=&{1\over 2}\int dx\left[{\pi^2\over 3}\tilde{\phi}^3(x)-
\int {1\over 2}\tilde{\phi}(x')\tilde{\phi}(x)(1-g(x+x'))dx'
+{1\over 4}(v'(x))^2\tilde{\phi}(x)\right.\nonumber\\
&-&\left.{1\over N}
{\tilde{\phi}(x')v'(x)\over (x-y)}+{2\over N}{\tilde{\phi}(x)\over x-y}
\int \!\!\!\!\!\!\! - \,\,\,\,\,\,
{\tilde{\phi}(x')\over x-x'}dx'+{1\over N^2}
{\tilde{\phi}(x)\over (x-y)^2}
\right].
\end{eqnarray}

\noindent From the expression for the effective potential derived in the last section,
it is easy to see it has a minimum value of zero. This minimum is achieved if 
$y$ is chosen to satisfy

\begin{equation}
\int \!\!\!\!\!\!\! - \,\,\,\,\,\, {\phi(x')\over y-x'}dx'={1\over 2}v'(y),
\end{equation}

\noindent
and we minimize $V_2$ with respect to $\tilde{\phi}(x)$.
The equation of motion following from this minimization is

\begin{eqnarray}
0&=&\pi^2\tilde{\phi}^2+{1\over 4}(v'(x))^2
-{1\over N}{v'(x)\over x-y}+{1\over N^2 (x-y)^2}
-\int dx'\tilde{\phi}(x')(1-gx'-gx)\nonumber\\
&+&{2\over N(x-y)}\int dx' {\tilde{\phi}(x')\over x-x'}
- {2\over N}\int dx' {\tilde{\phi}(x')\over (x'-y)(x-x')}.
\end{eqnarray}

\noindent
This equation is that of a Fermi surface but now with a deformation induced by the 
single eigenvalue instanton. Various deformations have been considered in the literature, e. g.
\cite{Aganagic:2003qj}.
This equation can be easily solved perturbatively in ${1\over N}$. 
The leading solution 
solves

\begin{equation}
0=\pi^2\tilde{\phi}_0^2+{1\over 4}(v'(x))^2
-\int dx'\tilde{\phi}_0(x')(1-gx'-gx).
\end{equation}

\noindent
Noting that $\int\tilde{\phi_0}(x)dx=1-{1\over N}$, we have

\begin{equation}0=\pi^2\tilde{\phi}_0^2+{1\over 4}(v'(x))^2
-(1-{1\over N})(1-gx)+c,
\end{equation}

\begin{equation}
c=g\int dx'\tilde{\phi}_0(x')x'\equiv c_0+{c_1\over N}.
\end{equation}

\noindent
The solution to first order in ${1\over N}$ solves

\begin{equation} 
0=2\pi^2\tilde{\phi}_0\tilde{\phi}_1
-{v'(x)\over x-y}
-\int dx'\tilde{\phi}_1(x')(1-gx'-gx)
+{2\over x-y}\int dx' {\tilde{\phi}_0(x')\over y-x'}.
\end{equation}

\noindent
Noting that $\int\phi_1 (x)dx=0,$ we have

\begin{equation}
0=2\pi^2\tilde{\phi}_0\tilde{\phi}_1
-{v'(x)\over x-y}
+\int dx'\tilde{\phi}_1(x')gx'
+{2\over x-y}\int dx' {\tilde{\phi}_0(x')\over y-x'}.
\end{equation}

\noindent
It is now a simple task to obtain

\begin{eqnarray}\tilde{\phi}_1&=&{1\over \pi}
{v'(x)-2\int dx' {\tilde{\phi}_0(x')\over y-x'}\over 2\pi(x-y)\tilde{\phi_0}}
-{\int dx'\tilde{\phi}_1(x')gx'\over 2\pi^2\tilde{\phi}_0}\nonumber\\
&=&{1\over \pi}
{v'(y)-2\int dx' {\tilde{\phi}_0(x')\over y-x'}\over 2\pi(x-y)\tilde{\phi_0}}
-{\int dx'\tilde{\phi}_1(x')gx'\over 2\pi^2\tilde{\phi}_0}+
{1\over \pi}
{v'(x)-v'(y)\over 2\pi(x-y)\tilde{\phi_0}}.
\end{eqnarray}

\noindent
For any polynomial potential, $v'(x)-v'(y)\propto x-y$ so that the last term in
$\phi_1(x)$ has no singularity at $x=y$. This result should be compared to the 
correction to the density obtained in section 2. We will be content to demonstrate
agreement between the singular term when $x\to y$ in both expressions. The singular 
term corresponds to the isolated eigenvalue, i.e. to the instanton. The non-singular
terms describe how the density of the other eigenvalues is distorted. The result 
for the singular term, from section 2, is

\begin{equation} 
{1\over \pi}{1\over y-x}{\sqrt{y-a_0}\sqrt{y-b_0}\over\sqrt{a_0-x}\sqrt{x-b_0}}.
\end{equation}

\noindent
The Fokker-Planck collective field theory gives

\begin{equation}
{1\over \pi}
{v'(y)-2\int dx' {\tilde{\phi}_0(x')\over y-x'}\over 2\pi(x-y)\tilde{\phi_0}}
={1\over \pi}{1\over y-x}{\sqrt{y-a_0}\sqrt{y-b_0}\over\sqrt{a_0-x}\sqrt{x-b_0}}
{f(y)\over f(x)},
\end{equation}

\noindent
for the same term. In this expression, $f(x)$ is a polynomial whose details are fixed
by the potential. The agreement between the above two expressions follows upon
noting that

\begin{eqnarray}
{1\over \pi}&&{1\over y-x}{\sqrt{y-a_0}\sqrt{y-b_0}\over\sqrt{a_0-x}\sqrt{x-b_0}}
{f(y)\over f(x)}\nonumber\\
={1\over \pi}&&{1\over y-x}{\sqrt{y-a_0}\sqrt{y-b_0}\over\sqrt{a_0-x}\sqrt{x-b_0}}
+{1\over \pi}{1\over y-x}{\sqrt{y-a_0}\sqrt{y-b_0}\over\sqrt{a_0-x}\sqrt{x-b_0}}
{f(y)-f(x)\over f(x)}.
\end{eqnarray}

\noindent
The second term on the right hand side is not singular as $x\to y$, so that the
equality is established. 

We will now consider the behaviour of $\tilde{\phi}_1$ in the double scaling limit.
>From section 2, we know that

\begin{equation}
\tilde{\phi}_1 =-{1\over \pi}{1\over y-x}
{\sqrt{y-a_0}\sqrt{y-b_0}\over\sqrt{a_0-x}\sqrt{x-b_0}}
+{1\over \pi}{1\over\sqrt{a_0-x}\sqrt{x-b_0}}.
\end{equation}

\noindent
Taking the double scaling limit as

\begin{eqnarray}
g=g_*(1-{3^{1/2}\over 16}\gamma^2 a^2 t),\\
x=x_*-\gamma az,\\
y=x_*+\gamma aw,
\end{eqnarray}

\noindent
we obtain

\begin{equation}
\rho_1 =-{1\over \pi\gamma}{a^{3/2}\over Na^{5/2}}{1\over w+z}
{\sqrt{w+\sqrt{t}}\over \sqrt{z-\sqrt{t}}}.
\end{equation}

\noindent
Note that this is exactly of the form expected for a $g_s$ effect and that the term which
is regular as $y\to x$ does not survive in this limit. This expression is again universal,
and independent of criticality, once the appropriate renormalized string coupling is identified.
It is in agreement with expressions obtained in conformal backgrounds ref\cite{KOPSS}.

\noindent
Consequently we have seen in this 
section that
the instanton solution can be generated in the (first) nonlocal form of the collective field 
equation.
Once the eigenvalue is separated a transition to the local form can be made. It takes the form 
of
a deformed Fermi surface with a deformation induced by the eigenvalue. At the original level 
of the
nonlocal equation however one has background independence, the instanton and the vacuum are 
both solutions of one and the same equation.

\section{Conclusions}

In this work we addressed the question whether D-instanton solutions can be obtained 
as solutions of a closed string field theory.  Due to the very special (stringy) nature 
of these objects, reflected in the fact that their action is given by $1/g$ as opposed to 
the standard $1/g^2$ of field theory, one could indeed expect that this is not possible. 
The studies of ref\cite{kawai} where the instanton effects were demonstrated at the level 
of matrices also pointed out some difficulties in obtaining these effects through loop or 
string equations. We believe that the instanton effects can be recovered in the continuum
collective field theory.  This required exploiting some subtle and nontrivial features of 
this field theory. We now list what these features were.  First, in evaluation of the free 
energy (or any correlator) one has to consider the presence of a nontrivial Jacobian which 
is seen to supply the crucial factor of $N$ toward the weight of an instanton contribution. 
Then at the level of field equations we emphasized the existence of two formally equivalent
equations. The first form was nonlocal while the second took the local form of a Fermi sea. 
It is in the first nonlocal version of the equation that the separation of a single eigenvalue
D-instanton is possible. Consequently one can state that the vacuum and the instanton are two 
(different) solutions of one and the same equation.  Considering the second local form of the
equation one sees a deformation (induced by the instanton). The collective field theory 
equations that we considered involve more degrees of freedom then the loop equations. In the 
sense of analytic continuation they involve both positively and negatively dressed loops.
We believe that this difference is responsible for the fact that (instanton) solutions are 
present in collective but not in loop equations. Furthermore one also had the effect of the
Jacobian which did not contribute to the form of the solution (in leading order) but did
contribute to the overall weight of the instanton contribution. It is also likely that the
D-instantons can also be described (probably more elegantly) in the framework of extended
open-closed field theory. After all, the eigenvalue deformations can be represented in terms 
of a quark integral. As a step toward this possible description of instantons we note that 
the nontrivial Jacobian that we emphasized very likely also posses an interpretation in terms 
of open string degrees of freedom.

$$ $$

\noindent
{\it Acknowledgements:} 
The work of R. de Mello Koch and J.P. Rodrigues is supported by NRF grant number 
Gun 2047219. The work of A. Jevicki is supported by DOE grant DE FG02/19ER40688
(Task A).

$$ $$

\appendix{Appendix A: BIPZ Solution for the eigenvalue density}

\noindent
In this appendix we summarize the solution to the integral equation

\begin{equation}
\int \!\!\!\!\!\!\! - \,\,\,\,\,\, {\phi(x')\over x-x'}dx'={1\over 2}\left(
x-gx^2+{2\over N(y-x)}
\right).
\end{equation}
Equations of this type have been investigated in connection with matrix theories of Penner 
type\cite{tan} and also theories of open and closed strings\cite{kazakov},\cite{kostov}.
The equation is solved  by extension of well known technique \cite{BIPZ}. For the case of a 
cubic potential this was given in  \cite{cicuta}. One begins by defining

\begin{equation} 
G(z)=\int \!\!\!\!\!\!\! - \,\,\,\,\,\, {\phi(x')\over z-x'}dx',
\end{equation}

\noindent
where $z$ is a complex number. From the properties of the principal value prescription,
it follows that

\begin{equation} 
G(x\pm i\epsilon)={1\over 2}\left(x-gx^2+{2\over N(y-x)}\right)\mp i\pi\phi (x)
\end{equation}

\noindent
where $x$ lies in the support of the density $\phi (x)$.
Analytic structure of $G(z)$ suggests the ansatz

\begin{equation}
G(z)=-g{z^2\over 2}+{z\over 2}-{1\over N(z-y)}+\left(
\bar{a}z+\bar{b}+{F\over y-z}
\right)\sqrt{z-a}\sqrt{z-b}.
\end{equation}

\noindent
The parameters $\bar{a}$, $\bar{b}$, $F$, $a$ and $b$ are determined by requiring that
(i) $G(z)$ does not have a pole at $z=y$ and (ii) that as $z\to\infty$, 
$G(z)\to {1\over z}$. The solution is

\begin{equation}
\phi (x)=-{1\over\pi}\left({g\over 2}x
-{1\over 2}+{g\over 4}(b+a)-
{1\over N\sqrt{y-a}\sqrt{y-b}}{1\over y-x}
\right)\sqrt{a-x}\sqrt{x-b}
\end{equation}

\noindent
with $a$ and $b$ solutions to the equations (in what follows
$s=a+b$ and $d=a-b$)

\begin{eqnarray}
{-16\over N\sqrt{y-a}\sqrt{y-b}}+s(2gs-4)+gd^2 &=&0,\\
-({s\over 2}-y){1\over N\sqrt{y-a}\sqrt{y-b}}
+{d^2\over 16}(1-gs)&=&1.
\end{eqnarray}

\noindent
We would like to compare this to the result of section 2. Towards this end
expand the density perturbatively in ${1\over N}$. To capture the next to leading
order, set

\begin{equation} 
a=a_0+{1\over N}a_1\qquad b=b_0+{1\over N} b_1\qquad s=s_0+{1\over N} s_1
\qquad d=d_0+{1\over N} d_1 ,
\end{equation}

\noindent
and

\begin{equation}
\phi (x)=\phi_0+{1\over N}\phi_1 .
\end{equation}

\noindent
We find

\begin{equation}
\phi_0(x)={1\over 2\pi}\left(1-gx
-{g\over 2}(b_0+a_0)\right)\sqrt{a_0-x}\sqrt{x-b_0},
\end{equation}

\begin{equation} 
s_0(2gs_0-4)+gd_0^2 =0,\qquad {d_0^2\over 16}(1-gs_0)=1,
\end{equation}

\noindent
for the leading order and

\begin{eqnarray}
\phi_1(x)&=&{1\over 2\pi}\sqrt{a_0-x}\sqrt{x-b_0}\Big[
{\sqrt{y-a_0}\sqrt{y-b_0}\over (a_0-b_0)}
\left({2\over (y-b_0)(x-b_0)}
-{2\over (y-a_0)(x-a_0)}\right)\nonumber\\
&-&{gs_1\over 2}
+\left(1-gx-{g\over 2}(a_0+b_0)\right)\left({a_1\over 2(a_0-x)}
+{b_1\over 2(b_0-x)}\right)\Big]
\nonumber\\
&-&{1\over \pi}
{\sqrt{y-a_0}\sqrt{y-b_0}\over \sqrt{a_0-x}\sqrt{x-b_0}(y-x)}
\end{eqnarray}

\begin{equation}
-{16\over\sqrt{y-a_0}\sqrt{y-b_0}}+4s_1 (gs_0-1)+2gd_0 d_1 =0,
\end{equation}

\begin{equation}
-\left({s_0\over 2}-y\right){1\over \sqrt{y-a_0}\sqrt{y-b_0}}
+{d_0 d_1\over 8}(1-gs_0)-g{d_0^2\over 16}s_1=0,
\end{equation}

\noindent
for the subleading order. 
The coefficient of the ${1\over y-x}$ pole is in perfect 
agreement with the result of section 2.

\end{document}